\documentclass[10pt,twocolumn,english,conference]{IEEEtran}
\usepackage{amsmath,epsfig}
\usepackage{amssymb}
\usepackage{amsfonts}

\usepackage{array,dcolumn}
\usepackage[ruled,vlined,linesnumbered]{algorithm2e}
\usepackage{psfrag}
\usepackage{graphicx}
\usepackage{amsmath}
\usepackage{amsfonts}
\usepackage{amssymb}
\usepackage{accents}
\usepackage{booktabs}
\usepackage{cite}
\usepackage{subfigure}
\usepackage[all]{xy}
\usepackage{dsfont}
\usepackage{url}
\usepackage{xcolor}

\begin{document}

\title{Reengineering GSM/GPRS Towards a Dedicated Network for Massive Smart Metering}

\author{\IEEEauthorblockN{Germ\'an Corrales Madue\~no, \v Cedomir Stefanovi\' c,   Petar Popovski}
\IEEEauthorblockA{\\Department of Electronic Systems, Aalborg University, Denmark \\
Email: \{gco,cs,petarp\}@es.aau.dk}}

\maketitle

\begin{abstract}
	GSM is a synonym for a major success in wireless technology, achieving  widespread use and high technology maturity.
	However, its future is questionable, as many stakeholders indicate that the GSM spectrum should be refarmed for LTE.
	On the other hand, the advent of smart grid and the ubiquity of smart meters will require reliable, long-lived wide area connections. 
	This motivates to investigate the potential of GSM to be evolved into a dedicated network for smart metering. 
	We introduce simple mechanisms to reengineer the access control in GSM. 
	The result is a system that offers excellent support for smart metering, as well as the other massive machine-to-machine traffic patterns that are envisioned in 3GPP.
\end{abstract}

\section{Introduction}

	Smart metering is a key machine-to-machine (M2M) application, where meters autonomously report usage and alarm information to the grid.
	It requires sending of low amounts of data from a very high number of meters, differing significantly from the high data rate requirements in human-oriented services.
	Cellular networks are mainly optimized for the latter, and the amendments required for the effective support of M2M services, including smart grid services, came only recently in focus of the standardization \cite{standardization}.
	As a general observation, the key technical problem in enabling M2M communications is not how to increase the overall system data rate, but how to distribute it efficiently to many terminals. 

	The clear dominance of 2G based solutions in M2M \cite{ABI} motivates us to investigate if and how GSM networks can be evolved into efficient smart metering networks.
	Our findings shows that, with a suitable reengineering, GSM networks can support a surprisingly massive M2M devices at even a single frequency channel. 
	This suggest that it is viable to keep one or few GSM channels for M2M operation in the coming years, and thus take advantage of its maturity, low cost and worldwide availability.

	The paper is centered on the assessment of the stages of the GSM radio access, which are random access (RACH), access granted (AGCH) and data transmission (DATA).
	Specifically, we elaborate the operation and the limitations of the GSM access and propose enhancements of the AGCH and DATA stages, with the aim of supporting large number of smart meters per cell.
	We also present a model of GSM radio access that considers the interstage dependencies, and show that the adopted 3GPP methodology, where the access stages are treated independently \cite{one,3GPP_comparison}, leads to unreliable results.
	Therefore, besides the main message of the paper, which endorses dedicated networks for smart metering based on GSM, the findings presented in the paper constitute an important contribution to the M2M-related 3GPP standardization~process. 
	
	The rest of the paper is organized as follows.
	Section~\ref{forecasted} describes the referent M2M traffic scenario.
	Section~\ref{system-operation} provides an overview of the GSM random access procedure.
	Section~\ref{reengineering} presents the proposed approach for the GSM access reengineering.   
	Section~\ref{simulation-methodology} describes the model used to asses the system performance and points out the shortcomings of the methodology used by 3GPP.
	Section~\ref{Results} presents the results.
	Section~\ref{Conclusions} concludes the paper.
	
\section{Traffic Models and General Requirements for Machine Type Communications}\label{forecasted}

	To assess the potential of GSM as smart metering network, we consider a referent M2M traffic scenario from \cite{IEEE802.16p0005}.
	The scenario includes the traffic originating both from smart meters and from commercial and home devices.
	The devices are deployed in a sub-urban GSM cell with a radius of 1000~m and three sectors.\footnote{Compared to the other GSM cell type - urban GSM cells, sub-urban cells have an increased coverage zone, potentially serving more devices.}
	The traffic parameters considered are the average message size, the average message arrival rate, and the arrival distribution.
	Table~\ref{traffic} summarizes these parameters for the referent scenario, listing also the expected number of devices in the cell. 
	As presented, smart metering differs from other M2M applications, foreseeing two operational modes: \emph{periodic} and \emph{alarm} reporting.
	Periodic reporting is characterized by variable reporting rates and tolerance of report losses, i.e., if a report is not successfully received, the metering application waits for the next scheduled reception.
	The alarm reporting is event-triggered, where the allowed reception delay is up to $1$~minute and loss of reports is not tolerated \cite{smartgrid}.
	Further, the presented traffic patterns can be divided into two categories - synchronous and asynchronous.
	For \emph{asynchronous} traffic, the arrivals are not correlated across devices, and traffic patterns with uniform and Poisson distribution fall into this category.
	On the other hand, the traffic generated by alarm-reporting is \emph{synchronous}, as the alarm event is typically detected by a multitude of smart meters, thus correlating the initiation of their transmission requests. 

	\begin{table*}[t]
		\begin{tabular}{p{130px}>{\centering}p{130px}>{\centering}p{70px}>{\centering}p{60px}p{40px}}
			\toprule
				Appliances/ Devices 				& Arrival rate [$\text{s}^{-1}$]   									& Average Message Size [byte] 	& Number of Devices  &  Distribution \\
			\midrule
				Smart Meters - Periodic Reporting					& 1.67e-2, 3.33e-3, 1.11e-3, 2.78e-4, 4.63e-5, 2.32e-5, 1.16e-5		& $<$1000		& 13941  				  &  Poisson\\
				Smart Meters - Alarm Reporting						& /													& $<$1000						& 13941  				  &  Beta(3,4) \\				
				Home Security System	 			& 1.67e-3		  	  												& 20 							& 3098	  				  &  Poisson  \\
	  		Elderly Sensor Devices 					& 1.67e-2		   	   	  											& 128 							& 310	  				  &  Poisson  \\ 				
				Credit Machine in Grocery 			& 8.3e-3		   		 										    & 24 							& 72					  &  Poisson  \\
	  		Credit Machine in Shop 					& 5.56e-4 	   		  												& 24 						    & 1100					  &  Poisson  \\ 
	 			Roadway Signs 						& 3.33e-2 		   		  											& 1 							& 2963					  &  Uniform  \\ 
				Traffic Lights	 					& 1.67e-2 		   		  											& 1 							& 360					  &  Uniform  \\
				Traffic Sensors 					& 1.67e-2 		   		 											& 1 							& 360					  &  Poisson  \\
				Movie Rental Machines				& 1.16e-4 	   		  												& 152							& 36					  &  Poisson  \\
			\bottomrule
		\end{tabular} 
		\caption{Traffic parameters for home and city commercial M2M devices in a sub-urban area for a 1000~m-radius cell, where 3 smart meters per home are considered \cite{smartgrid,IEEE802.16p0005, 3GPP_Smart }.}
		\label{traffic}
	\end{table*}	
		
	It can be inferred from Table~\ref{traffic} that a GSM base station should handle communication scenarios with high density of devices, small payload sizes, and sporadic transmissions.
	One of the main limitations of GSM access, which prevents its application in such scenarios, is the limited allocation granularity of the access resources. This type of bottleneck is valid for any TDMA-based system, as described next. 

	\subsection{The Resource Allocation Granularity Problem}

		\begin{figure}[t]
  			\centering
   			\includegraphics[width=0.9\columnwidth]{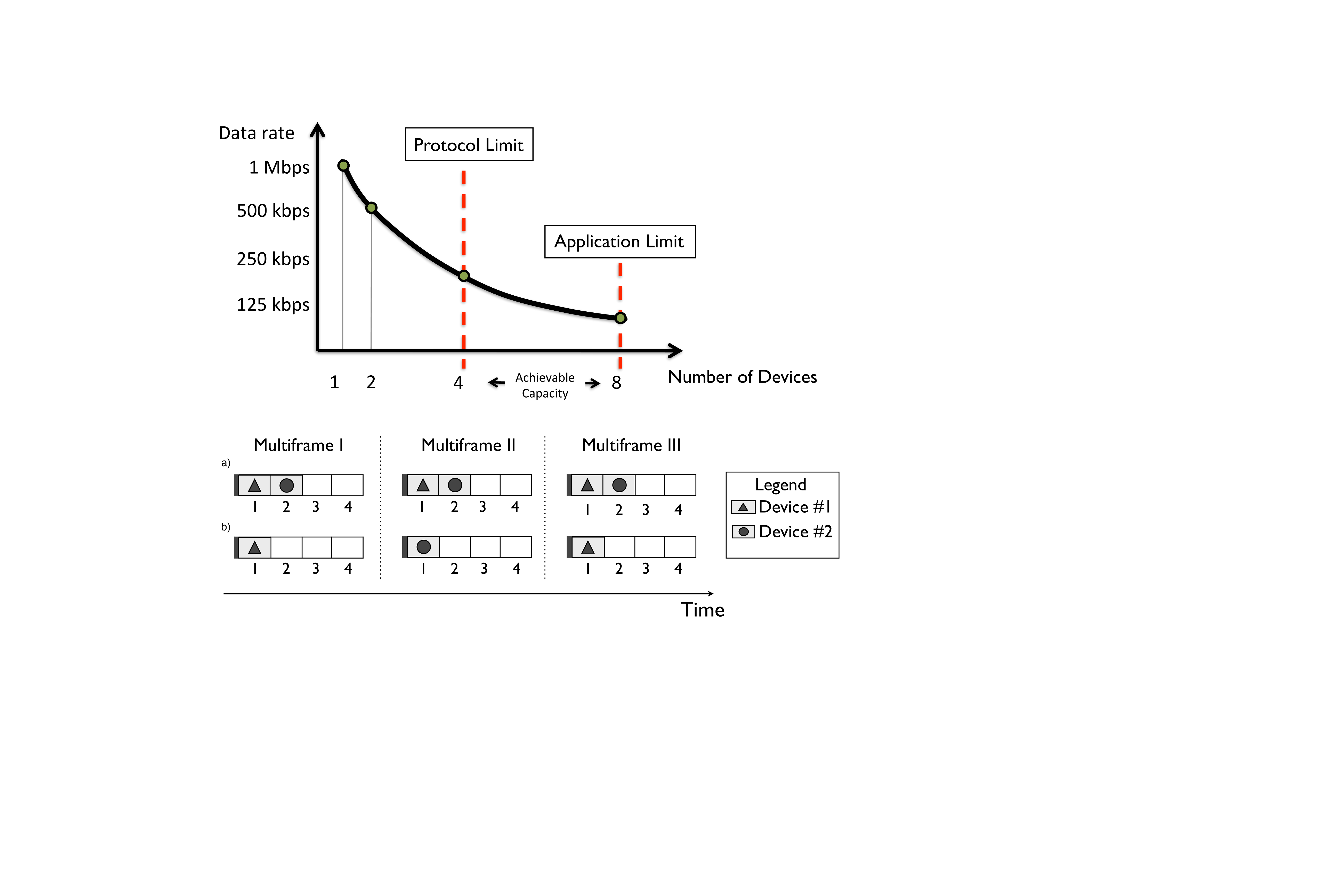}
				\caption{Ideal system in which the bandwidth is shared among the multiplexed devices. The protocol operation limits the number of devices: a) allocation capacity is limited to 4 devices, b) allocation capacity is improved to 8 devices.}\label{rateVSdevices}
		\end{figure}
		
		Ideally, the granularity of resource allocation in a TDMA system should be dictated by the application requirements and tuned to the application with the lowest demand of resources.
		However, in practice, TDMA allocation does not work this way, as the resource granularity is fixed by the allocation mechanism.
		Consider a toy example of a TDMA system with transmission data rate of $1$~Mbps, in which transmissions are organized in frames, and each frame consists of $4$ slots, Fig.~\ref{rateVSdevices}.
		It is assumed that the minimum number of slots that the system can allocate is a single slot per device per frame.
		Thus, if only one device is present in the system, it is allocated all $4$ slots and thus gets a total data rate of $1$~Mbps.
		If there are $4$ devices, each gets one slot and a data rate of $250$~Kbps, as illustrated in Fig.~\ref{rateVSdevices}.
		However, if devices require only $125$~Kbps, the system has the capacity to potentially accommodate $8$ of them, but, due to the constraints of the allocation mechanism, more than $4$ simultaneous connections cannot be supported at the same time.
		A straightforward approach is to re-allocate 4 devices in every new frame, but this introduces extra control traffic. 
		
		On the other hand, the fact that overall allocation is deterministic, i.e., each device should have the opportunity to transmit once in every two frames, could be exploited to design a more efficient allocation method. 
		An approach proposed in this paper is to \emph{logically} extend the allocation space by using frame numbers, known to all devices.
		Then both device \#1 and device \#2 can be allocated the same slot, but \#1 accesses it only in odd- and \#2 only in even-numbered frames, as shown in Fig~\ref{rateVSdevices}b).
		Although rather simple, the example illustrates both the limitations that are present in a real system such as GSM, as elaborated in the next section, and the main idea behind the method for their mitigation, elaborated in Section~\ref{reengineering}.
	
\section{GSM System Operation and Limitations}\label{system-operation}

\subsection{GSM Access Mechanism}\label{random_access}

\begin{figure}[t]
 			\centering
    			\includegraphics[width=\columnwidth]{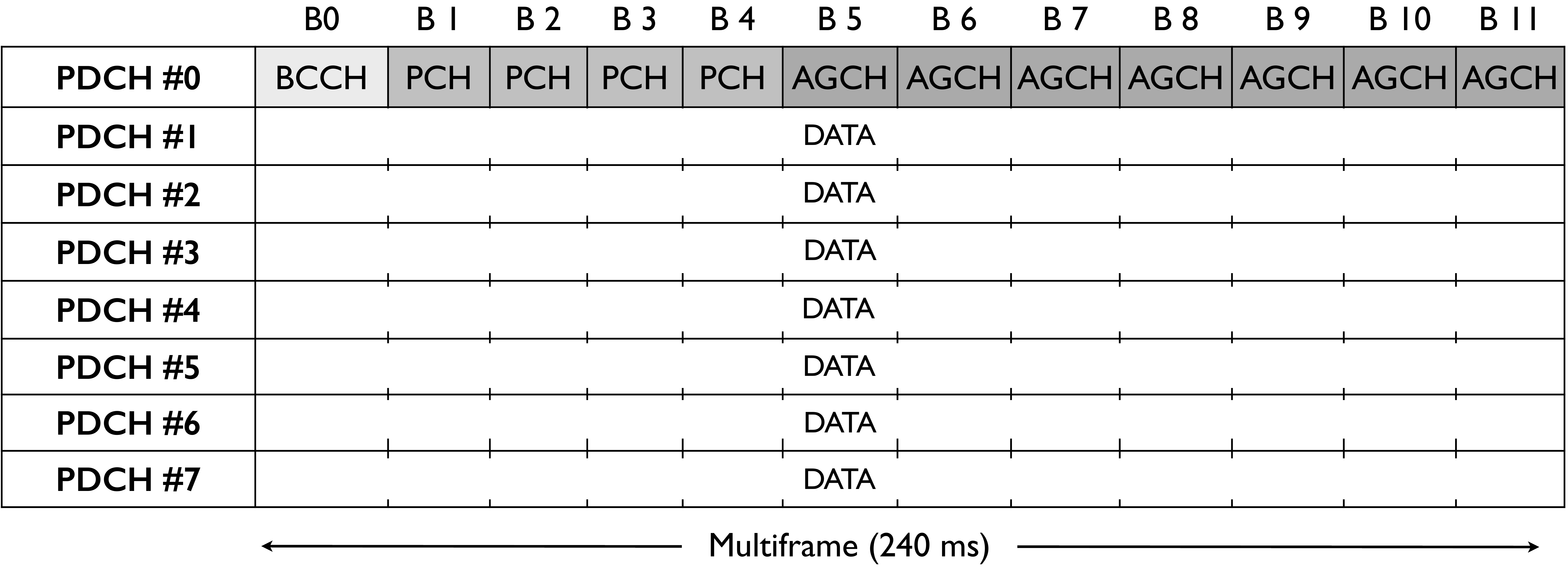}
				\caption{The structure of the downlink multiframe: the signaling channel PDCH~\#0 consists of one block dedicated to broadcast channel (BCCH), four blocks for paging channel (PCH) and the remaining blocks are is devoted to access granted channel (AGCH); the remaining PDCHs are devoted to data. }\label{multiframe}
		\end{figure}

	The access in GSM is TDMA-based, where both the uplink and the downlink are organized in multiframes with duration of $240$~ms.
	The multiframe structure, in its usual 2D representation, is shown in Fig.~\ref{multiframe}.
	A multiframe consists of 12 blocks, where a block is composed of four TDMA frames and a TDMA frame contains 8 time slots.
	Each time slot is interpreted as a separate TDMA channel, referred to as Packet Data Channel (PDCH).
	A PDCH can be dedicated either to signaling or data; in a typical configuration, PDCH~$\#0$ is for signaling and the remaining 7~PDCHs for data transmissions.
	
	The procedure to establish a connection with the base station (BS) consists of three stages, denoted as RACH, AGCH and DATA stage.
	The first stage is slotted ALOHA-based random access, in which devices contend for data resources.
	Specifically, a device with a pending data transmission first waits between 0 and $T-1$ frames, where the actual waiting time is randomly selected  and $T$ is a parameter broadcasted by the BS.
	In the next step, the device transmits its resource request in the random access channel (RACH) that is logically defined in the uplink PDCH~\#0, and waits for the response during the next $S$ frames, where $S$ is another parameter broadcasted by the BS.
	As the devices choose the transmission instants in an independent and uncoordinated manner, their resource requests can potentially collide in the RACH, as in any slotted ALOHA-based scheme.

	The BS responds to a resource request only if: (1) the request was successfully received, (2) there are available downlink resources to send the response, and (3) the requested uplink data resources are also available.
	The response is transmitted in the access granted channel (AGCH) that is logically defined in the downlink PDCH~\#0, and it has to be delivered before the waiting time of $S$ frames expires at the device side.
	If no AGCH message is received from the BS during this period, the device repeats the procedure until either a response is received or a maximum of $M$ retries is reached, where $M$ is also broadcasted by the BS.

	The response AGCH message assigns to a device an uplink PDCH and an uplink state flag (USF).
	USF is an identifier that controls the pending uplink data transmission; specifically, a device is allowed to transmit in block $k+1$ of the assigned uplink PDCH, if its USF was announced in block $k$ in the same downlink PDCH.

\subsection{GSM Access Limitations}\label{summary}
	
	As outlined, in order to establish a data connection, a device has to go through three different stages: RACH, AGCH and DATA.
	The amount of resources (i.e., blocks of the multiframe) devoted to each stage should be scaled to accommodate both the number of connection attempts and the expected traffic volume.
	Otherwise, if there are no sufficient resources in any of the stages, the ultimate result will be a situation in which a large number of devices are retransmitting their resource requests.
	This in turn will cause the RACH channel to collapse due to collisions of the retransmissions.
	Thus, to assure that the operation of the access network is not compromised, 3GPP recommends a blocking probability below 2\% in each of the stages \cite{3GPP_comparison}.
	The focus of the further text is on the limitations present in AGCH and DATA stages and the methods how to overcome them~by~MAC~protocol~reengineering.\footnote{As already outlined, the RACH stage is based on slotted ALOHA, whose limitations and improvements have been in the research focus for a long time.}
		
	A typical configuration foresees that 7 out of 11 blocks of the signaling PDCH are dedicated to AGCH, providing capacity to send approximately 30 AGCH messages per second \cite{Pavia}.
	However, as shown in Section~\ref{Results}, this configuration is inadequate for the referent scenario with a high density of low-rate devices, as a single cell cannot deliver enough AGCH messages within the time required to grant all resource requests.
	This limitation can be partially solved by dedicating more PDCHs to signaling, but this comes at the expense of the resources available for the DATA stage.

	In the DATA stage, the granularity of data resources is limited by the USF allocation mechanism.
	Specifically, USF is only 3 bits long and the value $000$ is reserved to indicate that the upcoming uplink block can be used for RACH contention. Thus, a maximum of 7 devices per PDCH can be multiplexed simultaneously.
	Therefore, in a single frequency configuration with 7~PDCHs devoted to data, the limit is 49 data connections per uplink multiframe, which is insufficient to support the traffic patterns of the referent scenario.
		
\section{Reengineering GSM for Massive Smart Metering}\label{reengineering}

	The solution for the AGCH bottleneck, proposed by Qualcomm \cite{qualcomm_solution}, is based on the observation that, in principle, M2M devices share the same capabilities and are likely to request the same type of service.
	Therefore, most of the AGCH message content can be directed to multiple devices requesting data resources.	
	Specifically, \cite{qualcomm_solution} foresees that four consecutive RACH requests could be granted with a single AGCH message, i.e., the capacity of the AGCH stage is increased four times in comparison to the legacy system.
	
	However, once the AGCH bottleneck is removed, the USF limitation becomes even more pronounced, as shown in Section~\ref{Results}.
	In the following text we present a method to remove the USF bottleneck, which in combination with the AGCH solution results with the improved GSM access scalability.

	The modification of the USF allocation mechanism is based on the method partially presented in \cite{us}. 
	The main conception behind it is that the \emph{validity range of USF is reinterpreted}, allowing for accommodation of a substantially increased number of active connections.
	In the following text we present the extended version of the solution, called \emph{expanded USF (eUSF)}, designed both for the periodic and alarm reporting. 

	In case of periodic reporting, the allocated USF does not hold anymore for an entire PDCH in all multiframes during which the connection is active, as prescribed by the GSM standard, but it is valid only for a specific set of blocks within a PDCH during $X$ consecutive multiframes, reoccurring periodically every $M$ multiframes. 
	In this way, several active devices reuse the same USF in the same PDCH, as they are multiplexed in the non-overlapping blocks and/or multiframes.
	The price to pay for this increased multiplexing capability is decreased data rate per active connection; however, the majority of M2M services require low data rates (see Table~\ref{traffic}), thereby rendering the proposed solution highly relevant. 
	
	\begin{figure}[t]
 			\centering
    			\includegraphics[width=\columnwidth]{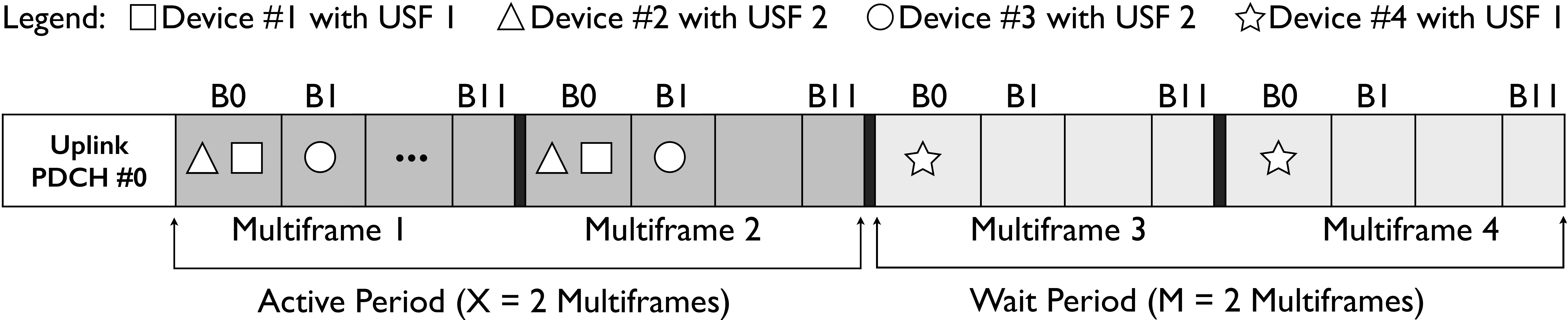}
				\caption{An example of expanded USF allocation, where the parameters $X$ and $M$ are set to 2: four devices are allocated in a single uplink PDCH with only three USFs.}\label{eUSF}
		\end{figure}
			
	We illustrate the eUSF method through an example depicted in Fig.~\ref{eUSF}. 
	There are four devices in the example, while the parameters $X$ and $M$ are set to 2, i.e., the allocated USFs' validity pattern consists of periods of 2 ``valid'' multiframes followed by 2 ``non-valid'' multiframes.
	Devices \#1, \#2 and \#3 arrive in multiframe 1; device \#1 is allocated USF1 that is valid in block 0, device \#2 is allocated USF2 in block 0, and device \#3 is allocated USF2 in block 1.
	Devices \#1 and \#2 are multiplexed in the same block of the multiframe using different USFs, which is supported by the standard GSM. 
	Devices \#2 and \#3, share the same USF in the same multiframe, but are multiplexed in different blocks - this way of operation is not supported in the standard GSM. 
	Furthermore, device \#4 arrives in multiframe 3 and it is allocated USF1 in block 0. 
	Devices \#1 and \#4 now share the same USF in the same block of the same PDCH, yet their transmissions are multiplexed as they take place in different multiframes; this way of operation is also not supported in the standard GSM operation.	
	Finally, the standard GSM system has to allocate 4 USFs in order to accommodate 4 users in a single PDCH, whereas in the above example this is done by using only 3 USFs.
	If the number of devices requiring service increases, a standard GSM system would rapidly run out of available USFs, as it is limited by its inflexible allocation method.
	On the other hand, the scalability of the eUSF solution is superior and limited only by the required data rates.
	Another advantage of the proposed method is that, once a device has been allocated a USF, this allocation can in principle last as long as required.
	Hence, the allocated devices do not have to go through RACH and AGCH stages anymore, relaxing the operation of these stages as well.

	The solution for the case of alarm traffic is similar, the main difference is that allocated USF is valid only for the specific set of blocks in the next $X$ consecutive multiframes, rather than being periodic.
	Once the device sends the report, the data connection is terminated; on the next occasion, the device has to go through RACH and AGCH stages again.
		
	Finally, we briefly outline potential methods to identify the devices compatible with the proposed improvements, while allowing the remaining devices to operate as usual.
	A simple option is to split the RACH resources into two different groups, one reserved for the standard operation and the other reserved for M2M traffic from the devices compatible with the new solution.
	However, the drawback of this approach is that it may leave unused RACH resources.
	In an alternative method, the device informs the BS of its compatibility through values reserved for future use in the RACH request message\cite{TS44.060}.

\section{Capacity Analysis of GSM Access}\label{simulation-methodology}
	
	\begin{figure}
		\centering
			\includegraphics[width=\columnwidth]{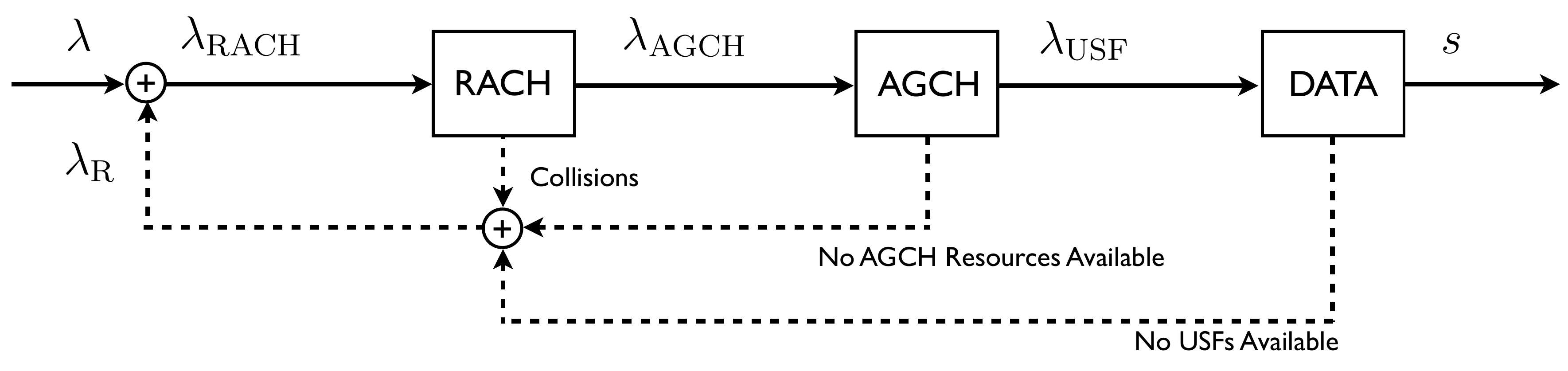}
			\caption{Illustration of the GSM access stages. The blocking probability for each stage should remain below 2\%.}\label{stages}
	\end{figure}
	
	Fig.~\ref{stages} depicts the stages of the GSM random access.
	The ``fresh'' traffic arrivals are represented by arrival rate $\lambda$, which takes into account all the arrivals of newly generated traffic of all the smart meters in the cell. 
	In the ideal case, all the devices with a pending transmission go through all the stages without being blocked, and re-enter the idle mode once the transmission is finished.	
	However, due to collisions in the RACH, lack of AGCH messages and lack of free USFs, the actual arrival rate at each stage is decreased by the success probability of the previous stage:
	\begin{align}
	\label{eq:stages}
		\lambda_\mathrm{AGCH} = \lambda_\mathrm{RACH} \cdot P_\mathrm{RACH}, \\
		\lambda_\mathrm{USF}  = \lambda_\mathrm{AGCH} \cdot P_\mathrm{AGCH},
	\end{align}
	where $P_\mathrm{RACH}$ is the probability that RACH request is successfully received and $P_\mathrm{AGCH}$ is the probability that a successful RACH request is being granted timely access by the BS.
	In other words, in general it holds:
	\begin{align}
		\lambda_\mathrm{RACH} \geq \lambda_\mathrm{AGCH} \geq \lambda_\mathrm{USF} \label{lambdas}.
	\end{align}
	The devices that are blocked in any of the stages retransmit their requests, generating an additional traffic represented by the arrival rate $\lambda_R$.
	Thus, the total arrival rate present in the RACH $\lambda_\mathrm{RACH}$ equals the sum of retransmission $\lambda_\mathrm{R}$ and ``fresh'' traffic arrivals $\lambda$.
	
	3GPP has studied the capacity of GSM access to serve M2M traffic in \cite{3GPP_comparison,3GPP_AGCH,3GPP_USF}.
	The adopted methodology assumes that the arrival rates to the AGCH stage $\lambda_\mathrm{AGCH}$ and the DATA stage $\lambda_\mathrm{USF}$ follow the same Poisson process that is present at the RACH stage, and does not take into account the impact of the retransmissions.
	I.e., 3GPP methodology assumes that: 
	\begin{align}
	\label{eq:wrong}
		\lambda = \lambda_\mathrm{RACH} = \lambda_\mathrm{AGCH} = \lambda_\mathrm{USF}.
	\end{align}
	However, this is rather approximate, as demonstrated next.
	
\begin{figure}[t]
		\centering
		\includegraphics[width=0.76\columnwidth]{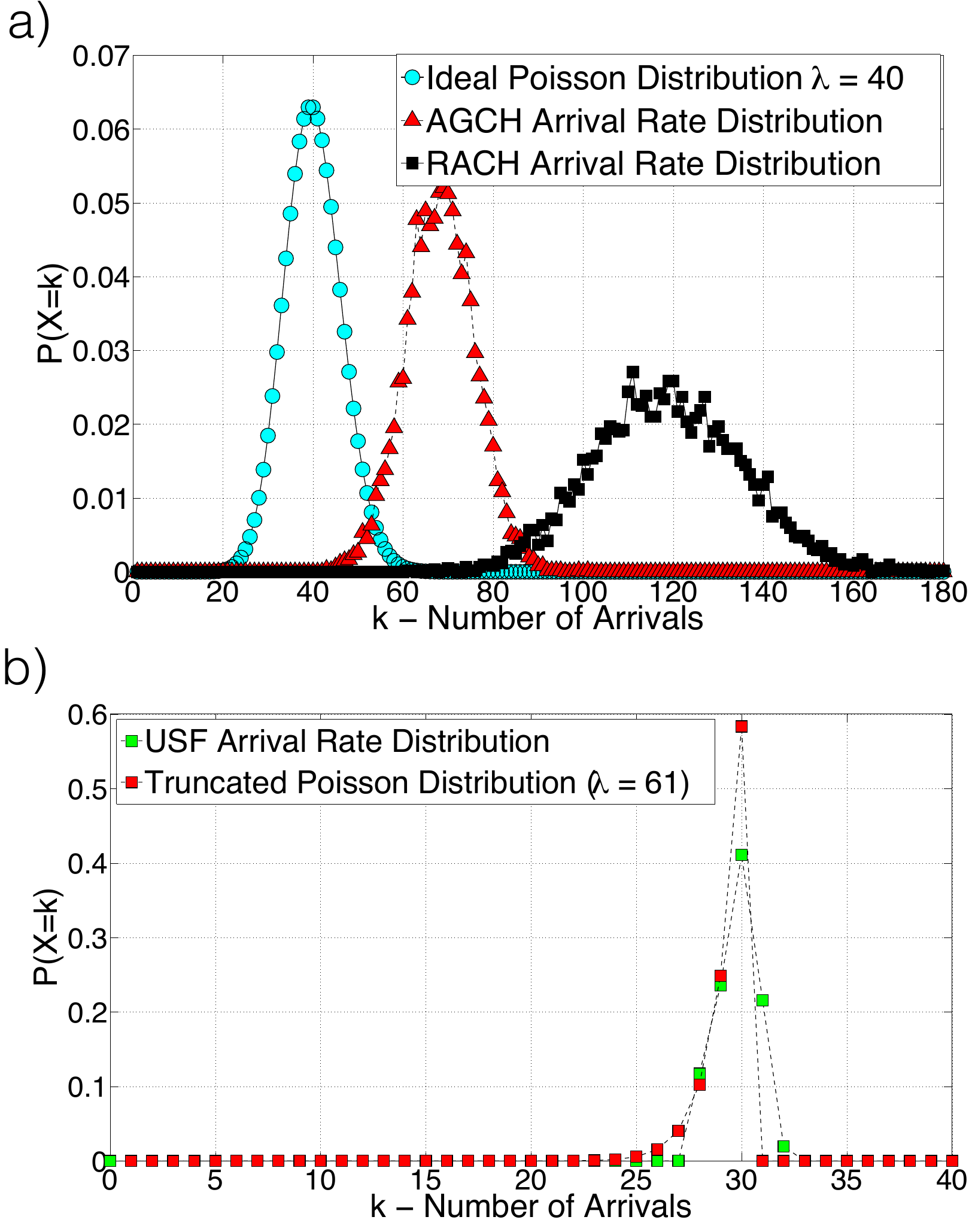}
		\caption{Arrival rate distributions of the smart meters traffic at a) RACH and AGCH stages and b) DATA stage, when the newly generated traffic is Poisson distributed with $\lambda = 40$.}\label{distributions}
	\end{figure}

	Fig.~\ref{distributions}a) presents the distribution of the traffic arrivals of smart meters for the RACH and AGCH stages, when the newly generated traffic is Poisson distributed with a mean of $\lambda=40$~arrival/s.\footnote{The rest of the parameters used for this study, i.e, the payload size and the GSM coding scheme, are described in Section~\ref{Results}.}
	It can be observed that the impact of the retransmissions is particularly pronounced, as the mean arrival rate to the RACH stage is $\lambda_\mathrm{RACH}\approx 120$~arrival/s, i.e., it is three times larger.
	Also, the mean AGCH arrival rate in this case is approximately equal to $\lambda_\mathrm{AGCH}=61$~arrival/s, far below $\lambda_\mathrm{RACH}$.
	The corresponding traffic arrival distribution to the DATA stage is depicted in Fig.~\ref{distributions}b).
	Obviously, the DATA stage arrival rate is limited to (approximately) $30$~arrival/s - which is a direct observation of the AGCH stage limitation, see Section~\ref{summary}.\footnote{We note that in \cite{3GPP_AGCH} a maximum of 38 AGCH/s is assumed; nevertheless, this fact does not impact our conclusions.}
	Our findings indicate that arrival process to the DATA stage can be described by a truncated Poisson distribution, obtained by truncating the Poisson distribution with mean $\lambda_\mathrm{AGCH}$ at the limit established by the AGCH capacity bottleneck; this is also demonstrated in Fig.~\ref{distributions}b).\footnote{The small deviation between analytical and simulation results is due to fact that a second is not multiple of multiframe duration.}
	
	Further, according to the 3GPP study \cite{3GPP_comparison}, the USF bottleneck is identified to be the most restrictive one with respect to M2M communications.
	Although this conclusion seems expected, due to the limited number of USF identifiers that can be allocated per multiframe, our findings indicate that the AGCH bottleneck is actually the most restrictive.
	The main reason is that the AGCH channel cannot grant enough number of devices per second in order to reach the maximum number of active connections supported in the DATA stage.
	Moreover, we again emphasize that in \cite{3GPP_comparison} the USF capacity is characterized using an input Poisson process with mean arrival rate $\lambda_\mathrm{USF} = \lambda_\mathrm{RACH}$, while the input arrival process should be actually modeled with a truncated Poisson distribution, as demonstrated above.
	
	In the next section, we demonstrate the potential of the proposed bottleneck solutions.
	Due to the interdependencies among GSM access stages, we adopt a numerical approach when investigating the performance of the GSM access, both for the legacy and the reengineered system.

\section{Results}
\label{Results}
	
	\begin{figure*}[t]
		\centering
		\includegraphics[width=0.95\textwidth]{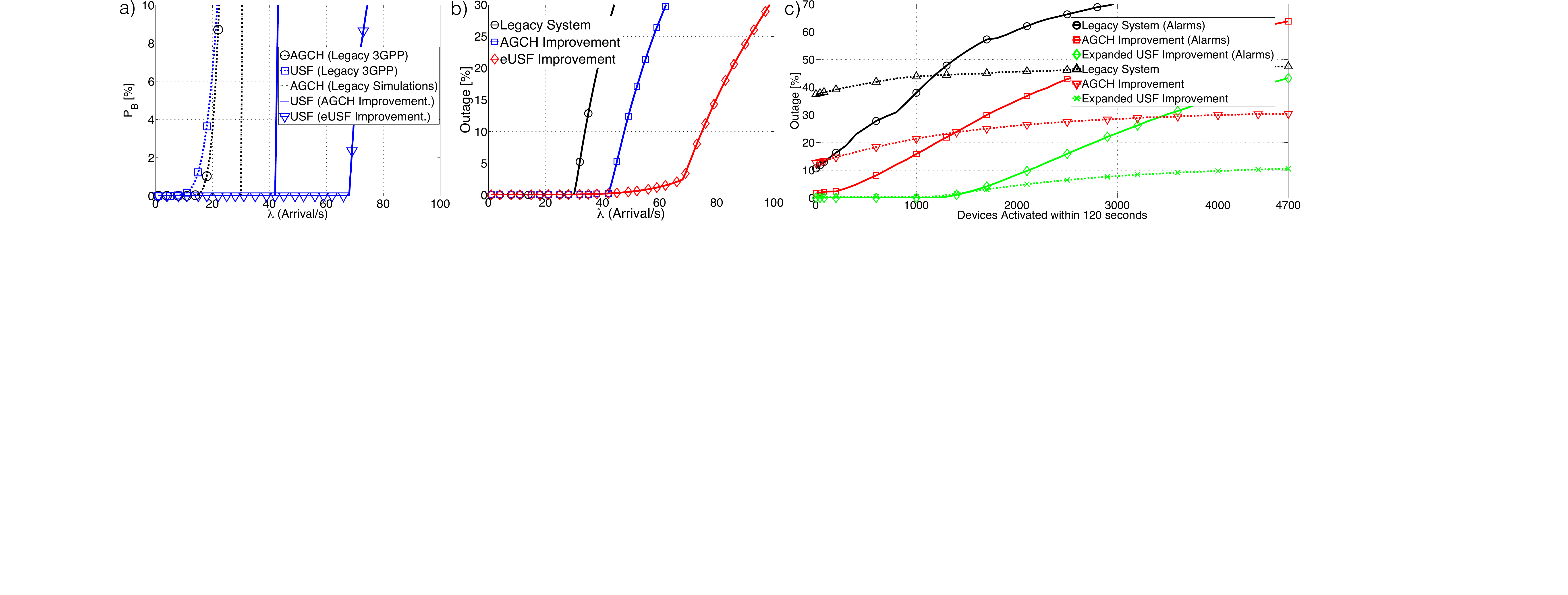}
			\caption{a) Blocking probabilities for AGCH and DATA stages, obtained by 3GPP model and simulations for asynchronous traffic. b) Outage of the legacy system, the system with AGCH improvement, and the system with AGCH and eUSF improvements for asynchronous traffic. c) Outage for asynchronous and synchronous reporting. The total asynchronous traffic is 42 arrival/s and the synchronously reporting smart meters are activated within 120 seconds.}\label{results}
		\end{figure*}
	
	The results presented in this section assume that all the asynchronously reporting devices have a uniform payload size of 152~bytes and that the most robust physical coding scheme (CS1) is used.
	This coding scheme represents the worst case scenario with a payload of 22~bytes per block.
	As shown in Table~\ref{traffic}, the selected application payload size is the upper limit for the presented M2M applications, excluding smart metering. 
	In case of smart metering, the M2M-related capacity analysis of 3GPP foresees payloads of 100, 500 and 1000~bytes\cite{3GPP_USF}.
	Due to space constraints, we show the results only for the synchronously reporting smart meters with a payload size of 100 bytes, and note that similar improvements can be observed for other payload sizes.
	We also assume that there are no channel-induced errors in the uplink/downlink transmissions, as our study is concentrated on the characterization of the access mechanism.
	For the simulations we have used a typical GSM configuration, where the RACH parameters are $T=20$, $S=105$ frames and maximum $M=4$ RACH retransmissions\cite{Pavia}.
	The performance parameters of interest are blocking probabilities of the AGCH and DATA stages, and the outage, defined as the fraction of the accessing devices that have reached the maximum number of connection attempts $M$ without establishing data connection.

	Fig.~\ref{results}a) compares the blocking probabilities $P_B$ of the AGCH and DATA stages as function of the system input arrival rate $\lambda$, when only asynchronous traffic is present in the cell (e.g., no alarm events.).
	The results are presented for (1) the legacy GSM access, (2) the reengineered GSM access where only AGCH solution is applied, denoted as AGCH improvement, and (3) the reengineered GSM access where both the AGCH and eUSF solutions are applied, denoted as eUSF improvement, both for the 3GPP methodology and numerical analysis.
	Obviously, the results obtained by the 3GPP methodology deviate both qualitatively and quantitatively from the ones obtained by the approach that takes into account interstage dependences.
	Specifically, the simulations show that the 3GPP blocking probabilities are overestimated and that, in the legacy system, the AGCH stage bottleneck is actually reached before the USF bottleneck.
	Also, when the AGCH bottleneck is removed, the effects of the USF bottleneck become obvious, and the eUSF solution demonstrates its full potential.
	The remaining blocking probabilities are not shown, as their values are zero for all the considered arrival rates.
	
	Fig.~\ref{results}b) compares the outage performance in the same scenario, obtained by numerical analysis.
	Obviously, superior results are achieved when AGCH and eUSF solutions are combined.
	E.g., a single-frequency GSM cell can support up to 70 arrivals per second with an outage below 2\%, which is an improvement of 133\% in comparison to the legacy system.
	In a 3-sector cell, this translates to approximately 13000 smart meters reporting every 5 minutes in addition to the expected asynchronous M2M traffic shown in Table~\ref{traffic}.
	
	Fig.~\ref{results}c) presents the analysis of the proposed AGCH and USF improvements for the traffic scenario presented in Section~\ref{forecasted}.
	The total expected arrival rate of the asynchronous traffic, modeled by uniform and Poisson distributions, is 42~arrival/s.
	The behavior of the synchronously reporting smart meters is modeled by a Beta distribution; we assume that the activation period, a central parameter of the Beta distribution, is set to 120~seconds.\footnote{The similar performance was observed for shorter activation periods.}
	Our goal is to investigate the GSM access performance as the number of synchronously reporting meters is increasing - i.e, we are interested to assess the behavior of the system when it is ``stressed'' by synchronously initiated resource requests.
	As pointed out by \cite{Pavia}, the RACH performance is severely affected by synchronous arrivals; the RACH limitations are outside the scope of this work and therefore we assume that the resource requests for each traffic type are transmitted in separate RACH channels.
	The comparison of the results reveals that the combination of the AGCH and eUSF solutions outperforms by far the legacy system and the system where only AGCH solution is implemented.
	Specifically, the combined AGCH/eUSF solution can roughly support up to 1500 synchronously reporting devices with an outage that is below rather demanding 0.1\%, as required for massively deployed sensors \cite{METISD2.1}.
	In case of more relaxed upper bounds on outage levels, the number of supported synchronously reporting meters rises, e.g., it is 2300 for the outage below 10\%.
	We note that neither the legacy system nor the AGCH only solution can assure outage level below 0.1\%.
		
\section{Conclusions}\label{Conclusions}			

	In this paper we have presented a concept to transform GSM into a dedicated network for massive smart metering, based on a simple redefinition of the access mechanisms.
	The demonstrated reengineering principles have been applied to AGCH and DATA stages of the GSM access, significantly boosting the performance in comparison to the legacy system.
	Results show that a single 3-sectorial cell can provide service up to 13000 smart meters reporting every 5 minutes in addition to the expected M2M traffic per cell.
	Also, up to 1500 synchronously reporting meters can be supported for rather demanding outage levels of 0.1\%.
The proposed changes are incurred only at the access control layer, leaving the physical interfaces intact, which a highly desirable feature in practice.
	
	Another important conclusion presented in the paper is that 3GPP modeling methodology of the GSM access is not valid, and that the corresponding results are incorrect.

	Finally, we note that 3GPP access protocols from different generations share the same stages: random access, access granted and data transmission.
	Hence, the proposed approach can be extrapolated to other cellular systems; the investigation how to exploit the analogues principles for the improvement of the resource granularity in 4G is part of our ongoing work.

\section*{Acknowledgement}

	The research presented in this paper was supported by the Danish Council for Independent Research (Det Frie Forskningsr{\aa}d), grant no. 11-105159 ``Dependable Wireless Bits for Machine-to-Machine (M2M) Communications'' and grant no. DFF-4005-00281 ``Evolving wireless cellular systems for smart grid communications''.

\bibliographystyle{IEEETran}
\bibliography{bib}

\begin{thebibliography}{10}
\providecommand{\url}[1]{#1}
\csname url@samestyle\endcsname
\providecommand{\newblock}{\relax}
\providecommand{\bibinfo}[2]{#2}
\providecommand{\BIBentrySTDinterwordspacing}{\spaceskip=0pt\relax}
\providecommand{\BIBentryALTinterwordstretchfactor}{4}
\providecommand{\BIBentryALTinterwordspacing}{\spaceskip=\fontdimen2\font plus
\BIBentryALTinterwordstretchfactor\fontdimen3\font minus
  \fontdimen4\font\relax}
\providecommand{\BIBforeignlanguage}[2]{{%
\expandafter\ifx\csname l@#1\endcsname\relax
\typeout{** WARNING: IEEEtran.bst: No hyphenation pattern has been}%
\typeout{** loaded for the language `#1'. Using the pattern for}%
\typeout{** the default language instead.}%
\else
\language=\csname l@#1\endcsname
\fi
#2}}
\providecommand{\BIBdecl}{\relax}
\BIBdecl

\bibitem{standardization}
G.~Wu, S.~Talwar, K.~Johnsson, N.~Himayat, and K.~Johnson, ``{M2M}: From
  {m}obile to embedded {i}nternet,'' \emph{IEEE Commun. Mag.}, vol.~49, no.~4,
  pp. 36--43, 2011.

\bibitem{ABI}
{ABI Research}, ``{Cellular M2M connectivity services},'' Tech. Rep. {2012.}

\bibitem{one}
H.~Dahmouni, B.~Morin, and S.~Vaton, ``Performance modeling of {GSM/GPRS} cells
  with different radio resource allocation strategies,'' in \emph{IEEE WCNC,
  2005}, vol.~3, march 2005, pp. 1317 -- 1322 Vol. 3.

\bibitem{3GPP_comparison}
3GPP, ``{Bottleneck Capacity Comparison for MTC},'' {3rd Generation Partnership
  Project (3GPP)}, TSG GERAN~{\#46 GP-100895}, 2010.

\bibitem{IEEE802.16p0005}
{IEEE 802.16p}, ``{IEEE 802.16p Machine to Machine (M2M) Evaluation Methodology
  Document (EMD)},'' {IEEE 802.16 Broadband Wireless Access Working Group
  (802.16p)}, EMD {11/0005}, 2011.

\bibitem{smartgrid}
{3GPP}, ``{RACH intensity of Time Controlled Devices},'' {3rd Generation
  Partnership Project (3GPP)}, {TSG RAN WG2} {R2-102340}, 2010.

\bibitem{3GPP_Smart}
3GPP, ``{Smart Grid Traffic Behavior Discussion},'' {3rd Generation Partnership
  Project (3GPP)}, TR {R2-102340}, 2010.

\bibitem{Pavia}
R.~Paiva, R.~Vieira, and M.~Saily, ``{Random Access Capacity Evaluation with
  Synchronized {MTC} Users Over Wireless Networks},'' in \emph{IEEE VTC Spring,
  2011}.

\bibitem{qualcomm_solution}
3GPP, ``{Packet Channel Assignments to Multiple Mobile Device},'' {3rd
  Generation Partnership Project}, TSG GERAN~\#48~GP-101894,~2010.

\bibitem{us}
G.~{Corrales Madue\~no}, C.~Stefanovic, and P.~Popovski, ``{H}ow {M}any {S}mart
  {M}eters can be {D}eployed in a {GSM} cell?'' in \emph{IEEE ICC 2013 -
  Workshop Telecom R2S}, pp. 1283--1288.

\bibitem{TS44.060}
3GPP, ``{General Packet Radio Service (GPRS); Mobile Station (MS) - Base
  Station System (BSS) interface; Radio Link Control / Medium Access Control
  (RLC/MAC) protocol},'' {3rd Generation Partnership Project (3GPP)}, TS
  {44.060}, Sep. 2008.

\bibitem{3GPP_AGCH}
{3GPP}, ``{Downlink CCCH Capacity Evaluation for MTC},'' {3rd Generation
  Partnership Project (3GPP)}, TSG GERAN~{\#46 GP-100893}, 2010.

\bibitem{3GPP_USF}
3GPP, ``{USF Capacity Evaluation for MTC},'' {3rd Generation Partnership
  Project (3GPP)}, TSG GERAN~{\#46 GP-100894}, 2010.

\bibitem{METISD2.1}
{FP-7 METIS}, ``{Requirements and general design principles for new air
  interface},'' Deliverable {D2.1}, 2013.

\end{thebibliography}

\end{document}